\documentclass[aps]{revtex4}
\usepackage[T1]{fontenc}
\usepackage[latin1]{inputenc}
\usepackage{verbatim}
\usepackage{graphicx}

\begin{document}

\title{
Are Genetically Robust Regulatory Networks Dynamically Different from Random Ones? }

\author{Volkan Sevim$^{1,2}$}
\email{volkan.sevim@duke.edu}
\author{Per Arne Rikvold$^{1}$}
\email{rikvold@scs.fsu.edu}

\affiliation{$^{1}$ School of Computational Science, Center for Materials Research
and Technology, and Department of Physics, \\
Florida State University, Tallahassee, FL 32306-4120, USA\\
$^{2}$ Physics Department, Duke University, Durham, NC 27708, USA}
%
%

\newcommand{\eqref}[1]{(\ref{#1})}
\newcommand{\mx}{\mathbf}
\newcommand{\wmx}{\mathbf{W}}
\newcommand{\state}{\mathbf{s}}
\newcommand{\initst}{\state (0)}
\newcommand{\finst}{\state {^*}}
\newcommand{\sgn}{\mathrm{sgn}}
\newcommand{\prob}{\mathrm{P}}
\newcommand{\stilde} {\tilde s}
\newcommand{\seta} {{\cal A}_t}
\newcommand{\setb} {{\cal B}_t}

\begin{abstract}
We study a \index{genetic regulatory networks} genetic regulatory network 
model developed to demonstrate that 
\index{genetic robustness}genetic robustness 
can evolve through stabilizing
selection for optimal phenotypes. We report preliminary
results on whether such selection could result in a reorganization
of the state space of the system. For the chosen parameters, 
the evolution moves the
system slightly toward the more ordered part of the phase diagram. 
We also find that strong memory effects cause the 
\index{Derrida annealed approximation}Derrida annealed approximation 
to give erroneous predictions about the model's phase diagram. 
\end{abstract}
\maketitle

\section{Introduction}

Gene networks are extremely robust against genetic perturbations \cite{RobustnessReview:2003,WagnerBook}.
For example, systematic gene knock-out studies on yeast showed that
almost 40\% of genes on chromosome V have no detectable effects on
indicators like cell division rate \cite{Smith:1996}. Similar
studies on other organisms agree with these results \cite{RobustnessReview:2003,WagnerBook}. It is also known
that phenotypically, most species do not vary much, although they
experience a wide range of environmental and genetic perturbations.
This striking resilience makes one wonder about the origins, evolutionary
consequences, and mechanistic causes of genetic robustness. 

It has been proposed that genetic robustness evolved through 
stabilizing selection for a phenotypic optimum. Wagner showed that this
in fact could be true by modeling a developmental process within an
evolutionary scenario, in which the genetic interaction sequence
represents organismal development, and the equilibrium configuration of the gene network
represents the phenotype \cite{Wagner:1996}. His results show that the
genetic robustness of a population of model genetic regulatory networks
can increase through stabilizing selection for a particular equilibrium
configuration (phenotype) of each network. 

In this paper we investigate the effects of the 
\index{biological evolution}biological evolution of genetic
robustness on the dynamics of gene regulatory networks in general. 
In particular,
we want to answer the question whether the evolution process moves
the system to a different point in the phase diagram.
Below, we present some preliminary results. 

\section{Model}

We use a model by Wagner \cite{Wagner:1996}, which
has also been used by other researchers with minor modifications.
Each individual is represented by a regulatory gene
network consisting of $N$ genes. The expression level of each gene, $s_{i},$
has only two values, $+1$ or $-1$, expressed or not, respectively. The expression states change in time
according to regulatory interactions between the genes. The time evolution
of the system configuration represents an (organismal) developmental pathway. The discrete-time
dynamics are given by a set of nonlinear difference
equations representing a 
\index{random threshold network}random threshold network (RTN),
\begin{equation}
s_{i}(t+1)=\left\{ \begin{array}{cc}
\mathrm{\mathbf{\mathrm{sgn}}}\left(\sum_{j=1}^{N}w_{ij}s_{j}(t)\right), & \sum_{j=1}^{N}w_{ij}s_{j}(t)\neq0\\
s_{i}(t), & \sum_{j=1}^{N}w_{ij}s_{j}(t)=0\end{array}\right.
\;,
\label{eq:main}
\end{equation}
where sgn is the sign function and $w_{ij}$ is the strength
of the influence of gene $j$ on gene $i$. Nonzero elements
of the $N\times N$ matrix $\wmx$ are independent random numbers drawn from a standard normal
distribution. (The diagonal elements of \textbf{$\wmx$} are allowed
to be nonzero, corresponding to self-regulation.) The (mean) number
of nonzero elements in $\wmx$ is controlled by
the connectivity density, $c$, which is the probability that a $w_{ij}$ 
is nonzero. 

The dynamics given by Eq. \eqref{eq:main} can have a wide variety
of features. For a specified initial configuration $\initst$, the
system reaches either a fixed-point attractor or a limit cycle after
a transient period. The lengths of transients, number of attractors,
distribution of attractor lengths, etc. can differ from system to
system, depending on whether the dynamics are ordered, chaotic, or critical. 
The fitness of an individual is defined by whether
it can reach a developmental equilibrium, a certain fixed gene-expression
pattern, $\mathbf{s^{*}}$, in a {}``reasonable'' transient time. Further 
details of the model are explained in the next section.

\section{Monte Carlo Simulations\label{sec:Monte-carlo-simulations}}

\subsection{Generation and Robustness Assessment of Random Networks }

We studied populations of $\mathcal N=400$ random networks (founding individuals) 
with $N=10$. Each network was assigned a matrix $\wmx$ and an
initial configuration $\initst$. $\wmx$ was generated as follows. 
Each $w_{ij}$ was independently chosen to be nonzero with probability $c$.
If so, it was assigned a random number drawn from a standard gaussian
distribution, $N(\mu=0,\sigma=1)$. Then, each {}``gene'' of the
initial configuration, $s_{i}(0)$, was assigned either $-1$ or $+1$
at random, each with probability 1/2. 

After $\wmx$ and $\initst$ were created, the dynamics were started
and the network's stability was evaluated. If the system reached a
fixed point, $\finst,$ in $3N$ timesteps, then it was considered
stable and kept. Otherwise it was considered unstable, both $\wmx$
and $\initst$ were discarded, and the process was started over and repeated
until a stable network was generated.{} For each stable network, 
its fixed point, $\mathbf{s^{*}}$, was regarded
as the {}``optimal'' gene-expression state (phenotype) of the system. This is
the only modification we made to Wagner's model: he generated networks
with preassigned $\initst$ and $\mathbf{s^{*}}$,  whereas
we accept any $\finst$ as long as it can be reached within $3N$ timesteps
from $\initst$. 

After generating $\mathcal N=400$ individual stable networks, we 
analyzed their state-space structures
and evaluated their robustness as discussed in subsection
\ref{sub:Assesment-of-Epigenetic}.

\subsection{Evolution}

In order to generate a breed of more robust networks, a mutation-selection
process was simulated for all of the $\mathcal N=400$ random, stable networks as
follows. First, a clan of $\mathcal N^{\prime}=500$ identical copies
of each network was generated. For each clan, a four-step process
was performed for $T=400$ generations: 

\begin{enumerate}
\item Recombination: Each pair of the $N$ rows of consecutive matrices in the clan
were swapped with probability 1/2. Since the networks were already
shuffled in step 4 (see below), there was no need to pick random pairs.
\item Mutation: Each nonzero $w_{ij}$ was replaced with probability $1/(cN^{2})$
by a new random number drawn from the same standard gaussian distribution. Thus,
on average, one matrix element was changed per matrix per 
\index{Monte Carlo simulation}Monte Carlo step.
\item Fitness evaluation: 
Each network was run starting from the original initial condition, $\initst$. 
If the network reached a fixed point, $\mathbf{s}^{\dagger},$
within $3N$ timesteps, then its fitness,  
$\mathbf{\mathrm{f}(\mathbf{\mathbf{s}^{\dagger},\finst})=
\exp(-\mathrm{H}^{2}(\mathbf{\mathbf{s}^{\dagger},\finst})/\sigma_{\rm s}))}$, 
was calculated. 
Here $\mathrm{H}(\mathbf{\mathbf{s}^{\dagger},\finst})$, denotes
the normalized Hamming distance between 
$\mathbf{s}^{\dagger}$ and $\mathbf{\finst}$, and
$\sigma_{\rm s}$ denotes the strength of selection, $\finst$ is the optimal
gene-expression state, which is the final gene-expression state of
the original network that {}``founded'' the clan. We used $\sigma_{\rm s}=0.1$.
If the network could not reach a fixed point, then it was assigned
the minimum nonzero fitness value, $\exp(-1/\sigma_{\rm s}).$
\item Selection/Asexual Reproduction: The fitness of each network was normalized
to the fitness value of the most fit network in the clan. Then, a
network was chosen at random and 
duplicated into the descendant clan with probability equal 
to its normalized fitness. This process was repeated until
the size of the descendant clan reached $\mathcal N^{\prime}$. Then the old
clan was discarded, and the descendant clan was kept as the next generation.
Note that this process allows multiple copies (offspring) of the same
network to appear in the descendant clan, while some networks may
not make it to the next generation due to genetic drift.
\end{enumerate}
At the end of the $T=400$ generation selection, any unstable networks 
were removed from the evolved clan.


\subsection{Assessment of Robustness \label{sub:Assesment-of-Epigenetic}}

The mutational robustness of a network was assessed slightly differently
for random and evolved networks. For a random network, first,
one nonzero $w_{ij}$ was picked at random and replaced by a new random
number with the same standard gaussian distribution. Then, the dynamics
were started, and it was checked if the system reached the same equilibrium
state, $\finst$, within $3N$ timesteps. This process was repeated $5000c$
times using the original matrix (i.e., each mutated matrix was discarded
after its stability was evaluated). 
The robustness of the original network before
evolution was defined as the fraction of singly-mutated networks
that reached $\finst$. 

For the evolved networks, clan averages were used. For each of 
$\mathcal N^{\mathrm{opt}}\leq 400$
networks in a clan, robustness was assessed as described above with
one difference: the number of perturbations was reduced to 
$5000c/{\mathcal N}^{\mathrm{opt}}$
per network to keep the total number of perturbations used to estimate
robustness of networks before and after evolution approximately equal. 
The mean robustness of the those ${\mathcal N}^{\mathrm{opt}}$ networks
was taken as the robustness of the founder network after evolution.
Therefore, the robustness of a network after evolution is the mean
robustness of its descendant clan of stable networks.

\section{Results}

As Wagner pointed out, the stabilizing selection described above
increases the robustness of the model population against mutations
\cite{Wagner:1996}. 
However, it is not very clear what kind of a reorganization
in the state space occurs during the evolution. Also, it is not known whether this
robustness against mutations leads to robustness against environmental
perturbations. In this paper, we focus on the effects
of evolution in terms of moving the system to another point in the
phase diagram. In other words, we investigate whether the system becomes
more chaotic or more ordered after evolution. 

\begin{figure}[t]
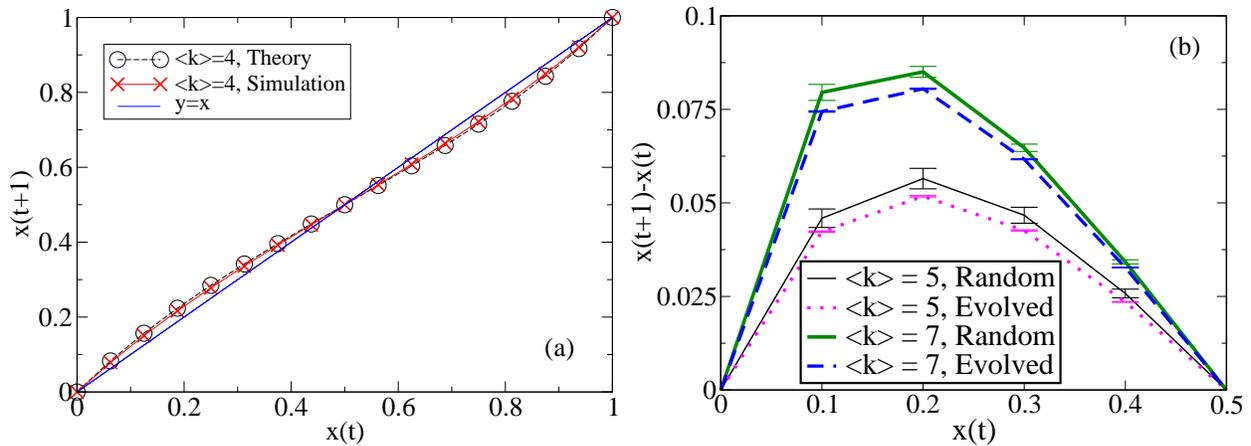

\vspace{0.4truecm}
\begin{center}
\includegraphics[width=.455\textwidth]{derrida-sim-and-theo--N=16-meank=4.eps}
\includegraphics[width=.455\textwidth]{random-and-evolved-maps-meank-5-and-7--N=10.eps}
\end{center}
\caption{
\label{fig:derridaplot} (a) $x(t+1)$ shown vs.\ $x(t)$ for $N=16$
and $\left\langle k\right\rangle =4$. The theory, 
Eq.~(\ref{eq:wagneroverlapmap}), is in good agreement
with the simulations. The deviations are due to the small
size of the simulated system as the theoretical calculation assumes
$N\gg\left\langle k\right\rangle $.
(b) Damage-spreading rate, $x(t+1)-x(t)$ vs.\ $x(t)$, 
for random and evolved networks with
$N=10$ and $\left\langle k\right\rangle =5$ and 7, showing the   
difference between the ``random'' and ``evolved'' curves. Only the first half 
of the curves are shown since $x(t+1)$ vs.\ $x(t)$
is point-symmetric about $(1/2,1/2)$.
The results were averaged over 10 random networks and all of
their evolved descendants ($\sim300$ evolved networks per random
network). The evolved curves for each $\left\langle k\right\rangle$ lie
very close to their ``random'' counterparts. However, they are
outside twice the error bar range of each other at most data points. 
}
\end{figure}
A standard method for studying damage spreading in systems such
as the one considered here 
is the Derrida annealed approximation \cite{Derrida:1986,Aldana:2003}, 
in which one calculates changes with time of the overlap of
two distinct states, ${{\mathbf s}(t)}$ and ${\mathbf \stilde}(t)$,
\begin{equation}
x(t)=\frac{1}{2N}\sum_{i=1}^{N}\left|s_{i}(t)+\stilde_{i}(t)\right|.
\end{equation}
The change of the overlap over one time step for $N\gg \langle k \rangle=Nc$ 
is given by
\begin{equation}
x(t+1) = n(0)x(t)+n(1)x(t)
+\sum_{k=2}^{\infty}n(k)\left[(x(t))^{k}
+\sum_{l=1}^{k-1}\Pi_{k}(l)\mathcal{P}(k,l)\right],
\label{eq:wagneroverlapmap}
\end{equation}
where the Poisson distribution $n(k)={\kappa^{k}\exp(-\kappa)}/{k!}$,
is the probability of finding a gene, $i$, with $k$ inputs, 
the binomial distribution
$\Pi_{k}(l)={k \choose l}\left(1-x(t)\right){}^{l}\left(x(t)\right){}^{k-l}$
is the probability of finding $k-l$ of these inputs in the overlapping 
parts of ${\mathbf s}(t)$ or ${\mathbf \stilde}(t)$, and
$\mathcal{P}(k,l)
=1-\frac{2}{\pi}{\mathrm {arctan}}\left(\sqrt{l/(k-l)}\right)$ (for $k>l$)
is the probability of the sum of $k-l$ matrix elements being larger than 
the sum of $l$ matrix 
elements, which are independent and $N(0,1)$ distributed. 
Here, $\kappa=\left\langle k\right\rangle $, 
the mean number of inputs per node. 

{}For most RTNs that have been studied so far
\cite{Derrida:1986,Aldana:2003},
Eq.~(\ref{eq:wagneroverlapmap}) can be iterated as a map to give the
full time evolution of the overlap. Changes in the fixed-point structure of
this map with changing $\langle k \rangle$ would then signify 
phase transitions of the system. 
As seen in Fig. \ref{fig:derridaplot}a, for $\left\langle k\right\rangle =4$, 
such a map would have a stable fixed point at $x=1/2$. 
One can also show that $\lim_{x(t)\rightarrow0^{+}}{\rm d}x(t+1)/{\rm d}x(t)>1$
for all $\left\langle k\right\rangle >0$ 
(this implies $\lim_{x(t)\rightarrow1^{-}}{\rm d}x(t+1)/{\rm d}x(t)>1$ 
and $\lim_{x(t)\rightarrow{1/2}}{\rm d}x(t+1)/{\rm d}x(t)<1$), 
and so it would seem that the 
system has no phase transition and always stays chaotic 
for nonzero $\langle k \rangle$.
However, simulations of damage spreading for longer times \cite{Sevim:2009} 
indicate that the system
studied here has strong memory effects due to the update rule for spins
with no inputs, given by the last line in Eq.~(\ref{eq:main}), which
retard the damage spreading \cite{DROS08}. In fact, like other RTNs
the system undergoes a phase transition near $\langle k \rangle \approx
2$ from a chaotic phase at larger $\langle k \rangle$ to an ordered
phase at smaller $\langle k \rangle$. The strong, retarding memory
effects mean that  
Eq.~(\ref{eq:wagneroverlapmap}) cannot be iterated as a map, and the
na{\"\i}ve prediction based on the Derrida annealed approximation is
erroneous. 

Despite its irrelevance for the long-time damage spreading, 
the damage-spreading rate shown in 
Fig.~\ref{fig:derridaplot}b 
properly describes the short-time dynamical character of the system.  
However, as Eq.~(\ref{eq:wagneroverlapmap})
assumes that the interaction constants, $w_{ij}$,
are statistically independent, it may not apply to evolved networks as we
do not know whether the selection process creates correlations between
the matrix elements. Nevertheless, we can still compute  
$x(t+1)$ as a function of $x(t)$ numerically 
to see if there is a change in the degree of chaoticity
(or order) of the dynamics. As seen in Fig. \ref{fig:derridaplot}b,
the damage-spreading rates for evolved networks are slightly 
(but statistically significantly) 
lower than for their random predecessors,
which are thus slightly more chaotic. 

To summarize, we have presented preliminary results on some general properties
of a popular RTN 
model of a gene regulatory network and on how the biological
evolution of genetic
robustness affects its dynamics \cite{Sevim:2008}. We have also shown that 
the update rule for spins without inputs leads to strong memory effects
that invalidate na{\"\i}ve iteration of the Derrida annealed
approximation as a map. 
The evolutionary process that improves the genetic
robustness of such networks has only a very small effect on their
dynamical properties: after evolution, the system moves slightly toward
the more ordered part of the phase diagram.


\section*{Acknowledgments}
We thank D. Balcan, B. Uzuno\u glu, and T.~F. Hansen for helpful discussions. 
This research was supported by U.S.\ National Science Foundation Grant 
Nos. DMR-0240078 and DMR-0444051, and by Florida State University through 
the School of Computational Science, the Center for Materials Research 
and Technology, and the National High Magnetic Field Laboratory.

%
%
%

%
\end{document}